## CONDENSED MATTER

# It's been a Weyl coming

Condensed-matter physics brings us quasiparticles that behave as massless fermions.

### B. Andrei Bernevig

"Mathematizing may well be a creative activity of man, like language or music"[1] — so said Hermann Weyl, the German physicist whose penchant for mathematical elegance prompted his prediction that a new particle would arise when the fermionic mass in the Dirac equation vanished[2]. Such a particle could carry charge but, unlike all known fermions, would be massless. During the course of his career, Weyl actually fell out of love with his prediction, largely because it implied the breaking of a particular symmetry, known as parity, which at the time was thought to be obeyed. More to the point, no such particle was observed during his lifetime.

After his death, the Weyl fermion was proposed to describe neutrinos, which are now known to have mass. For some time, it seemed that the Weyl fermion was destined to be just an abstract concept from another beautiful mind. That was until the Weyl fermion entered the realm of condensed-matter physics. For several years this field has been considered fertile ground for finding the Weyl fermion. Now, three papers in *Nature Physics*[3–5] have cemented earlier findings[6,7] to confirm the predictions[8,9] of Weyl physics in a family of nonmagnetic materials with broken inversion symmetry.

In condensed-matter physics, specifically in solid-state band structures, Weyl fermions appear when two electronic bands cross. The crossing point is called a Weyl node, away from which the bands disperse linearly in the lattice momentum, giving rise to a special kind of semimetal. The untrained eye would speculate that this is a very contrived situation, and that any small perturbation would open a gap that removes the node. However, in three spatial dimensions (that is, in bulk samples), it is impossible to open a gap because small perturbations merely shift the Weyl node in momentum k-space. The system therefore becomes a protected semimetal, wherein the true Weyl fermion is found.

This protection is simple to understand: two bands create a two-level system, which is describable by the 2 × 2 Pauli matrices. Because there are only three Pauli matrices, they can acquire the three lattice momenta as coefficients, and the Weyl node's k-space position is at k = 0. Any other perturbation introduced must couple to the Pauli matrices and hence can be re-interpreted as shifting the lattice momenta.

Perhaps the most remarkable way of understanding the protection of a Weyl node from opening a gap is through its so-called Berry phase. From the electron wavefunctions, one can build a quantity known as the Berry potential, which is the equivalent of a magnetic field in momentum space. The Weyl node is then interpreted as a monopole[10] in this 'magnetic field'. Every monopole has one end of a string attached to it, and in high-energy physics the other end of that string is taken to infinity. But in condensed-matter physics, the natural bandwidth cut-off means the string must end in another monopole (a Weyl node). Hence, in condensed matter, Weyl fermions appear in pairs — in high-energy physics, this is known as the fermion doubling 'problem'.

The Weyl resurgence in condensed matter physics came with the prediction[11] that the surface of a material, which contains Weyl fermions, would exhibit a new kind of surface state: an open Fermi arc (Fig. 1) that would connect two Weyl nodes and then continue on the opposite surface of the material. This arc would be readily observable in photoemission experiments[3–7], which map the electronic band structure of both surface and bulk states. The last thing needed for a fundamental experimental breakthrough was a realistic material proposal.

Although one proposal and experimental confirmation has been demonstrated in photonic systems[12,13], the particles in this study were bosonic in nature. Earlier this year, two groups[8,9] simultaneously predicted the existence of Weyl fermions in a series of compounds that include non-centrally symmetric transition metal monophosphides: TaAs, TaP, NbP and NbAs. The authors predicted the existence of 24 Weyl nodes in these materials, for which they computed monopole numbers and the Fermi arc patterns that would appear on specific surfaces. The clock was ticking for an experimental discovery.

Within a month and a half of these theoretical predictions, the first two papers[6,7] with preliminary data mapping the band structure of TaAs were concomitantly posted on the arXiv. These papers confirmed the theoretical prediction, but more data was needed to cement the notion that these materials are indeed Weyl semimetals.

Concomitantly with those experiments, three groups, whose papers are reported in *Nature Physics*[3–5], performed more complete photoemission experiments on both TaAs and another of the predicted compounds, NbAs, mapping their surface states and their bulk band structure. The bulk band structure resembles that of conical gapless bands merging to a point, which is the dispersion relation of a Weyl node.

Although the experimental resolution is not fine enough to identify whether the bulk band structure is indeed gapless (Weyl) or has a small gap (<1 meV), the combination of the bulk band structure and the mapping of the surface Fermi arcs (which come directly out of the projection of the Weyl nodes on the surface of the material) is very strong evidence in favour of the Weyl node. The projection of the bulk Weyl node onto the surface is shown to be the starting point of the Fermi arc. Although the exact structure of the Fermi arcs depends on the surface termination, the authors have identified several universal properties that can be associated only with the presence of Weyl points, as described below.

Because such photoemission experiments have momentum resolution, the authors provided evidence that in a closed loop of the surface Brillouin zone, there is an odd number of Fermi surfaces crossing the Fermi level; in other closed loops, there are no (or an even number of) Fermi surfaces crossings. Because bands are generically spin-split at the surface due to spin–orbit splitting and the lack of inversion symmetry, the crossing count described is consistent only with Fermi arcs — a trivial or nontrivial topological insulator or a trivial metal would not have this property. As theoretically predicted, these papers have proved that a new state of matter exists: the Weyl semimetal.

The future of the field is bright. Ultimately, we would like to find a unique transport signature of Weyl fermions. Although several studies have attempted to observe signatures of a phenomenon called the chiral anomaly[14–17], strictly speaking the chiral anomaly is a phenomenon present in a single Weyl cone; in condensed matter, Weyl points always come in pairs. More theoretical input is needed to understand and interpret those experiments, as well as to predict a transport property that is uniquely characteristic of the observed Weyl semimetals.

Now begins the search for the hydrogen atom of a Weyl semimetal, a magnetic material with only two Weyl nodes at the Fermi level. Moreover, additional terms can be added to the Weyl Hamiltonian to obtain a new type of Weyl fermion[18] that exhibits protected crossed Fermi surfaces and has different properties to those of the Weyl fermions predicted in 1929. Weyl neglected these terms because they break Lorentz invariance, which is a symmetry well-regarded in high-energy physics. But this symmetry is not present in condensed matter.

Elucidating the effects of disorder and looking at strongly interacting versions of the Weyl fermion, perhaps described by 3+1 spacetime dimensional conformal field theories, is also part of the future revolution of topological semimetals. The experimental finding of the Weyl semimetal will surely bring about renewed effort in this exciting area of physics.

B. Andrei Bernevig is an associate professor in the Department of Physics, Princeton University, Princeton, New Jersey 08544, USA. e-mail: bernevig@princeton.edu

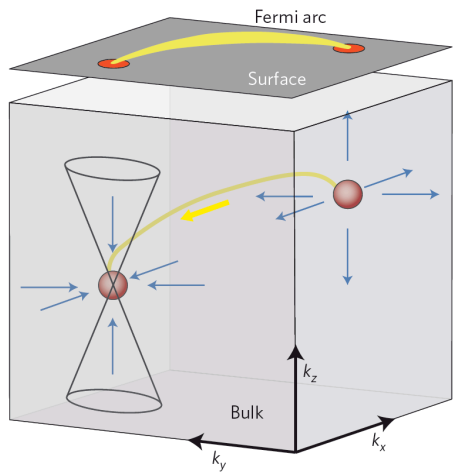

**Figure 1 | Weyl semimetals in momentum space.** Weyl semimetals in momentum space. Two Weyl nodes (red) act as monopoles, which have linear band dispersions (black) and are connected by a Dirac string (yellow). The top plane (grey) shows the two-dimensional projection, which has a Fermi arc (yellow) that connects the nodes and can be observed in photoemission experiments.